\documentclass[preprint, number]{elsarticle}

\makeatletter
\def\ps@pprintTitle{%
	\let\@oddhead\@empty
	\let\@evenhead\@empty
	\def\@oddfoot{}%
	\let\@evenfoot\@oddfoot}
\makeatother

\usepackage{lineno,hyperref}
\usepackage{amsmath}

\modulolinenumbers[5]
\journal{Information Sciences}

\biboptions{sort&compress}

\newcommand*\rfrac[2]{{}^{#1}\!/_{#2}} 
\begin{document}

\begin{frontmatter}
	
\title{Relativizing an incompressible number and an incompressible function through subrecursive extensions of Turing machines\tnoteref{mytitlenote}}

\tnotetext[mytitlenote]{This research was supported by the National Council for Scientific and Technological Development (CNPq), Brazil, as a PhD Fellowship at the Federal University of Rio de Janeiro (UFRJ), Rio de Janeiro, Brazil.}

\author[fsainst]{Felipe S. Abrah\~{a}o}

\address[fsainst]{Postdoctoral researcher at the National Laboratory for Scientific Computing (LNCC), Brazil }


	\begin{abstract}
		We show in this article that uncomputability is also a relative property of subrecursive classes built on a recursive relative incompressible function, which acts as a higher-order ``yardstick'' of irreducible information for the respective subrecursive class. We define the concept of a Turing submachine, and a recursive relative version for the Busy Beaver function and for the halting probability (or Chaitin's constant) $ \Omega $; respectively the Busy Beaver Plus (BBP) function and a time-bounded halting probability. Therefore, we prove that the computable BBP function defined on any Turing submachine is neither computable nor compressible by any program running on this submachine. In addition, we build a Turing submachine that can use lower approximations to its own time-bounded halting probability to calculate the values of its Busy Beaver Plus function, in the ``same'' manner that universal Turing machines use approximations to $ \Omega $ to calculate Busy Beaver values. Thus, the algorithmic information carried by the BBP function is relatively incompressible (and uncomputable) at the same time that it still is occasionally reached by submachines. We point that this phenomenon enriches the research on the relativization and simulation of uncomputability and irreducible information.
	\end{abstract}
	
	\begin{keyword}
		Algorithmic information \sep Subrecursion \sep Relative computability \sep Busy Beaver function \sep Halting probability \sep Time-bounded Turing machines
		\MSC[2010] 68Q30 \sep 68Q05 \sep 03D20
	\end{keyword}
	
\end{frontmatter}

\section[Introduction]{Introduction}

Let's remark some important concepts and definitions introduced in \cite{Abrahao2015, Abrahao2016}. Basically, we can define a \textbf{Turing submachine} as a Turing machine that always gives an output for any input, i.e. always halts. So, note that Turing submachine is just another terminology for \textbf{total Turing machines } \cite{Burgin2010, Kozen1997}. However, despite the fact that they would be just different names for the same object and can be used interchangeably, the expression ``total Turing machine'' might not immediately capture its relevant properties related to the present work. 

Remember that every computable total function (or total Turing machine) defines a subrecursive class which is a proper subclass of others subrecursive classes (and of the class of all recursive functions) \cite{Carnielli2008, Lewis1997, Sipser2006, Basu1970}. A subrecursive class is one defined by a proper subset of the set of all problems with Turing degree \boldmath ${0}$\unboldmath. Therefore, a \textbf{subcomputable\textit{ }}class of problems will be subrecursive, because it will never contain all recursive/computable problems. The term subrecursion is also used to characterize subrecursive hierarchies, as in Kleene and Grzegorczyk \cite{Hoyrup2016, Basu1970, Rose1984}, covering all primitive recursive functions. But for us, the prefix refers more specifically to the concept of subrecursive class \cite{Kozen1980, Hoyrup2016, Basu1970}.

For example, the total Turing machine may be a subsystem of another total machine which is capable of computing functions that are relatively uncomputable by the former. This very idea of being part of another non-reducibly more powerful machine is the core notion of the expression ``submachine'', conveying and bearing the ideas of hierarchies of subrecursive classes together with the powerful concept of Turing machines. Thus, the terminology Turing submachine emphasizes this property of total Turing machines being always able to be part of another proper and bigger machine. For more of this discussion, see \cite{Abrahao2016} and section \ref{submachines}. 

The main idea presented here is building, and rather proving, a system (a Turing machine) that can ``behave'' in relation to a subsystem (its Turing submachine) in the same way as a hypercomputer \cite{Copeland2002, Syropoulos2008} (or an oracle Turing machine \cite{Turing1939}) would behave in relation to a subsystem (in particular, a universal Turing machine). In fact, we will not emulate all --- which might be impossible \cite{Hoyrup2016, Kozen1980} --- the properties of a hypercomputer in relation to a computer, but focus on defining a function ${{BB}^+}_{{P^{**}}_T\circ P_T}(N)$ analogous to a Busy Beaver\textbf{\textit{ }}function (in Chaitin's first models \cite{Chaitin2012} in metabiology  \cite{Chaitin2013, Chaitin2014}, the function $BB\textrm{´}(N)$), and a recursive relative number\textit{ }${\mathrm{\Omega }}_{{P^{**}}_T\circ P_T}$ analogous to $\mathrm{\Omega }$, so that this function will behave in relation to the Turing submachine and to the number ${\mathrm{\Omega }}_{{P^{**}}_T\circ P_T}$ ``resembling'' the way the original Busy Beaver behaves in relation to a universal Turing machine and to $\mathrm{\Omega }$. 

In short,  ${{BB}^+}_{{P^{**}}_T\circ P_T}(N)$ must be relatively uncomputable and incompressible by any \textbf{subprogram} (a program running on a Turing submachine), the same --- except for a constant --- way the original Busy Beaver is incompressible by any program running on a universal Turing machine. We will call this phenomenon of relativization of the uncomputability of a hypercomputable function defined on Turing submachines as \textbf{recursive relative uncomputability, }or\textbf{ sub-uncomputability}, of a function (in our case, Busy Beaver function).

Also bear in mind that for the present purposes we do not want it to be ``overly'' relatively uncomputable, as a first order oracle machine is already capable of computing $BB'\textrm{´}(N)$.  We want the new relative Busy Beaver function to be ``as, but not more than'', uncomputable for a Turing submachine as is the Busy Beaver for a universal Turing machine. This way, it keeps playing a role of an outside measurement of irreducible complexity (or information) \cite{Chaitin2012, Zenil2011} while being occasionally reached by some subprograms. For this purpose, we will demonstrate that there are subprograms that calculate ${{BB}^+}_{{P^{**}}_T\circ P_T}\left(N\right)\ $values for lower approximations to\textbf{\textit{ }}${\mathrm{\Omega }}_{{P^{**}}_T\circ P_T}$, as is already known in the case of the Busy Beaver and of $\mathrm{\Omega }$. 

Before reaching the three theorems that this article aims, we start by defining some general conditions on a self-delimiting programming language for the universal Turing machine $U$. The definition of the Turing submachine $U_{{P^{**}}_T\circ P_T}$ playing the central role herein is intrinsically also dependent on the time-bounded halting probability\textbf{\textit{ }}${\mathrm{\Omega }}_{{P^{**}}_T\circ P_T}$\textit{ }and on the function ${{BB}^+}_{{P^{**}}_T\circ P_T}$, since the algorithm of the program ${P^{**}}_T\circ P_T$ is defined by  a successive and self-referential composition of other algorithms --- a construction that resembles a fixed-point (or diagonal) lemma \cite{Enderton2001, Regan1992} used to prove incompleteness results for example. Then, we prove that $U_{{P^{**}}_T\circ P_T}$ is indeed well-defined. It relies on the fact that bigger subprograms only depends on smaller subprograms to always halt, which enable us to build a proof by induction. The second theorem derives from the very construction of the program ${P^{**}}_T\circ P_T$ and from the former result. It proves the desired property of, for every $N$, function ${{BB}^+}_{{P^{**}}_T\circ P_T}$ being calculated by some subprograms through lower approximations to ${\mathrm{\Omega }}_{{P^{**}}_T\circ P_T}$ as inputs. And we finish by proving the uncomputability and incompressibility of the function Busy Beaver Plus by any program running on a Turing submachine. This proof, as previously presented in \cite{Abrahao2015, Abrahao2016}, is based on an argument analogous to the one in Chaitin's incompleteness theorem \cite{Chaitin1971}, which uses algorithmic complexity/information. Therefore, we conclude by showing that function ${{BB}^+}_{{P^{**}}_T\circ P_T}$ is sub-uncomputable by any program running on $U_{{P^{**}}_T\circ P_T}$, but still reachable by giving lower approximations to ${{\Omega}}_{{P^{**}}_T\circ P_T}$ as inputs to a class of programs running on $U_{{P^{**}}_T\circ P_T}$.

\section[Language L]{Language $L$} \label{introlanguage}
	
	It is important to us that the submachine $U_{{P^{**}}_T\circ P_T}$ can be programmable. Its language must be able to be used on any usual computer. Hence, the properties and rules of well-formation of the universal programming language for the chosen universal Turing machine $U$ upon which we will define submachines must be recursive. As one is invited to see below in definition \ref{language}, the funding conditions for the language $L$ are actually met by most self-delimiting programming languages for practical purpose. As we already have shown in \cite{Abrahao2016}, this leads us, along with theorem \ref{thm11.1}, to the conclusion that the phenomenon of sub-uncomputability is ubiquitous.
	
	Since we are trying to build a computer that can emulate some ``behaviors'' of  a first order oracle Turing machine with respect to a universal Turing machine, it is necessary that one can ``teach'' the machine to perform and recognize ``well-formed functionalizing concatenations'' within the language. That is, one needs a language that provides a direct way of symbolizing a program taking any given bit string as input (for example, a program $p$ that is, actually, program $p'\textrm{´}\ $taking program $p'\textrm{´}\textrm{´}$ as input), which makes this program act as a function. Note that this type of program is already used to demonstrate the\textit{ }halting problem, or to demonstrate that the Busy Beaver function is uncomputable for example. But the form it may assume is completely arbitrary, as a universal Turing machine --- note that it is not our case for the present purposes ---, in any case, will run it. 
	
	The desired condition is met by a recursively well-formed type of concatenation, which we will symbolize by ``$\circ $''. And we denote the optimal functionalizing concatenation, which is ``joining'' strings in the most compressed way possible, by ``$*$''.

\subsection[Definition of L]{Definition} \label{language}

We say a universal programming language $l$, defined on a universal Turing machine $U$, is \textbf{recursively functionalizable} if there is a program that, given any bit strings\textit{ }$P$ and $w\ $as inputs\textit{,} returns a bit string belonging to $l$ which will be denoted as $P\circ w$, whereby $U\left(P\circ w\right)\ $equals ``the result of the computation (on $U$) of program $P$ when $w$ is given as input''. In addition, there must be a program that determines whether or not a bit string is in form $P\circ w$ for any $P$ and $w$, and is capable of returning $P$ and $w$  separately. 

Analogously, it also must be true for the successive concatenation $P\circ w_1\circ \dots \circ w_k$, with program $P$ receiving $w_1$,{\dots}$w_k$ as inputs. 

So, the general definition of our language $L$ comes just below. 

Let $U$ be a universal Turing machine running language $L$, a universal language that is \textbf{binary, self-delimiting, recursive,} and \textbf{recursively functionalizable }such\textbf{ }that there are constants $\epsilon $, $C$ e $C'$, for every $P$, $w_1$,{\dots}$w_k$, where:

\[\left|w_i\right|\mathrm{<}\left|P\mathrm{\circ }w_{\mathrm{1}}\mathrm{\circ }\mathrm{\dots }\mathrm{\circ }w_k\right|,for\ i\mathrm{=1,2,\dots }or\mathrm{\ }k\]

\noindent and

\noindent 
\[\left|P\mathrm{\circ }w_{\mathrm{1}}\mathrm{\circ }\mathrm{\dots }\mathrm{\circ }w_k\right|\mathrm{\le }C\mathrm{\times }k\mathrm{+}\left|P\right|\mathrm{+}\left|w_{\mathrm{1}}\right|\mathrm{+|}w_{\mathrm{2}}\mathrm{|+\dots +|}w_k\mathrm{|}\]

\noindent and

\noindent 
\[H\left(N\right)\mathrm{\le }C\mathrm{'+}{{\mathrm{log}}_{\mathrm{2}} N\ }\mathrm{+}\left(\mathrm{1+}\epsilon \right){{\mathrm{log}}_{\mathrm{2}} \left({{\mathrm{log}}_{\mathrm{2}} N\ }\right)\ }\]

Let $H(w)$ denote the algorithmic complexity (or Solomonoff-Kolmogorov-Chaitin complexity) of the bit string $w$ defined on the machine $U$, i.e. $H(w)$ is the size of the shortest program $P$ in $L$ such that $U(P)=w$ (see definition 3.\ref{d3.b}). And let $H(N)$ denote the algorithmic complexity of the natural number $N$, that is, the size of the shortest program in $L$ that outputs\footnote{A representation (except for a trivial bijection) of $N$ in the language $L$ actually. } $N$ when run on the machine $U$ --- see definition 3.\ref{d3.d}.

\section[Definitions and Notations]{Some definitions and notations}

\begin{enumerate}[a)]
	
\item \textbf{ }$W$ is the set of all finite bit strings, where the computable enumeration of these bit\textit{ }strings has the form $l_1$, $l_2$, $l_3$, ..., $l_k$, ...\textit{ }

\noindent For practical purposes, a language may be adopted where $l_1=0$.

\item  Let $l$ be a language for a universal Turing machine $U$. Let $w\in l$. \label{d3.b}

\noindent Then, $U\left(w\right)$ denotes the result of the machine $U$ running the string $w$.

\item  Let\textbf{\textit{ }}$w\in W$.

\noindent Then, $|w|$ denotes the size or number of bits contained in \textit{w.}

\item\label{d3.d}  Let $N$ simply symbolize the corresponding program in language $L$ for the natural number $N$. For example, $P\circ N$ denotes program $P\circ w$ where $w$ is a representation of the natural number $N$ in the language $L$.

\item  If function $f$ is computable by program $P$, then $f$ may also be called function $P$.

\end{enumerate}

\section[Turing submachines]{Turing submachines} \label{submachines}

As discussed in \cite{Abrahao2016}, we follow the conventional understanding in which a computation that is a part of another computation may be called a subcomputation, and a machine that is a part of another machine may be called a submachine. For example, a Turing submachine can be a program or subroutine that the ``bigger'' Turing machine runs, always generating an output, while performing various other tasks. Note that it is true (a theorem) that for every total Turing machine there is another Turing machine that completely simulates and contains the former total Turing machine, in a manner that the computations of the latter contains the computations of the former. So, a system can be taken as a Turing machine, and a subsystem can be taken as a Turing submachine.

In fact, we are using a stronger notion of subsystem based upon this conventional notion: a subsystem must be only able do what the system knows, determines and delimits. This way, submachines will only be those machines for which there is another non-reducibly ``bigger'' machine that can decide, at least, what is the output of the former and whether there is an output at all. In particular, this condition proved to be necessary in order to build theoretical models for an open-ended\footnote{That is, an evolution of software through random algorithmic mutations in which programs ("organisms") always grow in algorithmic complexity (or gain more irreducible information) over time.} \cite{Hernandez-Orozco2016} evolution of programs in which the very environment, or ``Nature'', can be simulated in a computer \cite{Abrahao2015}. Note that every machine that falls under this definition always defines an equivalent \textbf{total Turing machine} (with a signed output corresponding to the case where the submachine does not halt); and every total Turing machine falls under this definition. 

Let $P_f$ be\textit{ }a program running on $U$ defined in$\ $language$\ L$, computing a total function (a function defined for all possible input values) $f$ such that $f:L\longrightarrow X\subseteq W$. The language $W$ does not need necessarily to be self-delimiting, and may be comprised of all bit strings of finite size, as long as they may be recursively enumerated in order, as $l_1$, $l_2$, $l_3$,\textit{... }For practical reasons, we will choose an enumeration where $l_1=0$.

A \textbf{``Turing submachine'', or total Turing machine,} ${U}/{f}$ is defined\textit{ }as a Turing machine in which, for every bit string $w$ in the language of $U$, ${U}/{f}\left(w\right)=U\left(P_f\circ w\right)$. 

This definition is quite general and transforms any total computable function into a Turing submachine. In fact, as said in the introduction, Turing submachines are just another name for total Turing machines. Anyway, submachines can always be subsystems of either abstract universal Turing machines or of powerful enough everyday computers (which are also some sort of total Turing machine, i.e. a universal Turing machine with limited resources).

Note that the class of all submachines is infinite, but not recursive. And when we talk of \textbf{subprograms }we refer to programs run on a Turing submachine. 

We will now use another concept of vital importance:\textbf{ computation time.} Similarly to time complexity, we will call $T$ a program that calculates how many steps or basic operations $U$ performs when running program $p$. Thus, if $U(p)$ does not halt, then $U(T*p)$ will not halt either, and vice versa. Herein, only submachines of a particular subclass will be dealt with: submachines defined by a computation time limited by a computable function, i.e. time-bounded Turing machines \cite{Homer2011}. In fact, both these and the more generic submachines defined above are equivalent in computational power --- although they might not be in definition. To demonstrate this, just note that if a program computes a total function, then there is a program that can compute the computation time of this first program. Therefore, for every computable total\textbf{\textit{ }}function, there is a submachine with bounded computation time capable of computing this function (but possibly other functions as well). The reverse directly follows from the definition of submachine. This holds even though the definition of a submachine may come from more arbitrary restrictive conditions (although still recursive) than just limiting the computation time by some upper bound gave by a computable function (e.g. a polynomial one).

Let $P_T$ be\textit{ }an arbitrary program that calculates a computation time for a given program $w$. That is, let\textit{ }$P_T$ be\textit{ }an arbitrary computable total\textbf{\textit{ }}function. Thus, there is a \textbf{Turing submachine} \boldmath ${{U}}_{{{P}}_{{T}}}$ \textbf{defined by the computation-time function }${{P}}_{{T}}$ \unboldmath.

We then define submachine $\rfrac{U}{{P_{SM}\circ P}_T}$ (which will be a program running on $U$ that computes a total\textbf{\textit{ }}function), where $P_{SM}$ is a program that receives $P_T$ and $w$ as inputs, runs $U(P_T\circ w)$ and returns:

\begin{enumerate}[(i)]
	\item  $l_1$, if $U(w)$ does not halt within computation time $\le U(P_T\circ w)$;
	
	\item  $l_{k\mathrm{+1}}$, if $U\mathrm{(}w\mathrm{)}$ halts within computation time $\mathrm{\le }U\mathrm{(}P_T\mathrm{\circ }w\mathrm{)}$ e $U\mathrm{(}w\mathrm{)=}l_k$;
\end{enumerate}

This program defines a Turing submachine that returns a known symbol (in our case, from definition 3.a, we have that $l_1=0$) when program $w$ does not halt in time $\le U\left(P_T\circ w\right)\ $or returns the same output\textit{ }(except for a trivial bijection) as $U\left(w\right)$ when the latter halts in time $\le U(P_T\circ w)$.

To be a Turing submachine, ${U}/{{P_{SM}\circ P}_T}$ must be defined for all inputs. This occurs because  $P_T$ is total by definition. In addition, as computation time $P_T$ becomes more increasing, the more submachine\textit{ }${U}/{{P_{SM}\circ P}_T}$ approaches the universality of $U$\textit{.}

Therefore, we will denote only as $U_{P_T}$ a Turing submachine\textit{ }${U}/{{P_{SM}\circ P}_T},$\textit{ }so that:

\[\mathrm{\forall }w\in L\ \mathrm{(\ }U_{P_T}\left(w\right)\mathrm{=}\rfrac{U}{{P_{SM}\mathrm{\circ }P}_T}\left(w\right)\mathrm{=}U\left(P_{SM}\mathrm{\circ }P_T\mathrm{\circ }w\right)\mathrm{)}\] 

\section[Function BBP]{Function ${{{BB}}^{{\mathrm{+}}}}_{{{P'}{\mathrm{\textrm{´}}}}_{{T}}}$} 

Let ${\mathrm{P'\textrm{´}}}_{\mathrm{T}}\mathrm{\ }$be a total function. Let us define function ${{BB}^+}_{{P'\textrm{´}}_T}(N)$, which we call \textbf{Busy Beaver Plus}, through the following recursive procedure:

\begin{enumerate}[(i)]

		\item  Generate a list of all outputs of $U_{{P'\textrm{´}}_T}\left(w\right)$ such that $\left|w\right|\le N$;
		
		\item  Take the largest number on that list;
		
		\item  Add 1;
		
		\item  Return that value.

\end{enumerate}

The name of this function refers to the Busy Beaver $BB\left(N\right)$ function and, consequently, it is no coincidence that the two have almost the same definition. If step (iii) is removed, it becomes exactly the Busy Beaver function for\textbf{\textit{ }}Turing submachines (denoted below as ${BB}_{{P'\textrm{´}}_T}(N)$). Thus:

\[{BB}^{\mathrm{+}}\left(N\right)\mathrm{=}BB\left(N\right)\mathrm{+1}\]

\noindent and

\[{{BB}^{\mathrm{+}}}_{{P'\mathrm{\textrm{´}}}_T}\left(N\right)\mathrm{=}{BB}_{{P'\mathrm{\textrm{´}}}_T}\left(N\right)\mathrm{+1}\]

But why use function ${BB}^+$ instead of $BB$? As we are dealing with Turing submachines and ${P'\textrm{´}}_T$ is arbitrary, it is possible that there is a program on $U_{{P'\textrm{´}}_T}$ with size $\le N$ such that computes the highest value returned by any other program on $U_{{P'\textrm{´}}_T}$ with size $\le N$. When dealing with a universal Turing machine $U$, this cannot occur (except for a constant). However, with submachines, it can. Thus, function ${{BB}^+}_{{P'\textrm{´}}_T}$ is triggered to assure it, in itself, is not relatively computable (and compressible) by any program on $U_{{P'\textrm{´}}_T}$, although it can be by a program on$\ U$. Note that, since ${P'\textrm{´}}_T$ is a program that computes a total\textbf{\textit{ }}function, then ${{BB}^{\mathrm{+}}}_{{P'\mathrm{\textrm{´}}}_T}\left(N\right)$ is computable.

The Busy Beaver contains the idea of the greatest output of any $\le N$ sized program; so the Busy Beaver Plus function contains the idea of surpassing (by $1$) any $\le N$ sized program. Respectively, the first gives us maximization, and the second, an ``almost'' minimal increment. No matter how rapidly increasing is the function ${P'\textrm{´}}_T$, the program on $U$ that computes ${{BB}^+}_{{P'\textrm{´}}_T}\left(N\right)$ simply bases itself on the $U_{{P'\textrm{´}}_T}$ outputs to overcome them by a minimum. No matter how powerful $U_{{P'\textrm{´}}_T}$ may be, ${{BB}^+}_{{P'\textrm{´}}_T}\left(N\right)$ will always be ``one step'' ahead of the best that any subprogram (i.e., any program $U_{{P'\textrm{´}}_T}$) can do.

It is worthy of note that, analogously to the Busy Beaver, the ${{BB}^+}_{{P'\textrm{´}}_T}\left(N\right)$ may be used to measure the algorithmic creativity or \textbf{sub-algorithmic complexity} \cite{Hoyrup2016} of the subprograms in relation to Turing submachine $U_{{P'\textrm{´}}_T}$, like a resource-bounded algorithmic complexity \cite{Li1997, Allender2011, Watanabe1992}. Why? By its very definition, if a subprogram generates an output $\ge {{BB}^+}_{{P'\textrm{´}}_T}\left(N\right)$, it must necessarily be of size $>N$. It needs to have more than $N$ bits of relatively incompressible information, i.e. more than $N$ bits of relative creativity. Further along, we will build a submachine $U_{{P^{\mathrm{**}}}_T\mathrm{\circ }P_T}$ that always runs a subprogram of size $\le 2N+C$ that calculates ${{BB}^+}_{{P^{\mathrm{**}}}_T\mathrm{\circ }P_T}\left(N\right)$, where $C\ $is a constant. Therefore, the sub-algorithmic complexity $H_{{P^{\mathrm{**}}}_T\mathrm{\circ }P_T}$ of function ${{BB}^+}_{{P^{\mathrm{**}}}_T\mathrm{\circ }P_T}\left(N\right)$ is such that

\[N<H_{{P^{\mathrm{**}}}_T\mathrm{\circ }P_T}\left({{BB}^+}_{{P^{\mathrm{**}}}_T\mathrm{\circ }P_T}\left(N\right)\right)\le 2N+C\]

\noindent In summary, as we will prove below in theorem \ref{thm10.2}, this allows submachine $U_{{P^{\mathrm{**}}}_T\mathrm{\circ }P_T}$ to occasionally grasp irreducible information ``from the outside'' through lower approximations to ${\mathrm{\Omega }}_{{P^{**}}_T\circ P_T}$.

Of course, one may always build a program that computes function ${{BB}^+}_{{P'\textrm{´}}_T}$, if  function ${P'\textrm{´}}_T$ is computable. This would allow a far smaller program than $N$ there to exist --- e.g., of size $\le C+{{log}_2 N\ }+(1+\epsilon ){{log}_2 ({{log}_2 N\ })\ }$ --- that will compute ${{BB}^+}_{{P'\textrm{´}}_T}\left(N\right)$. But that does not constitute a contradiction, because this program can never be a subprogram of $U_{{P'\textrm{´}}_T}$, in other words, it can never be a program that runs on the computation time determined by ${P'\textrm{´}}_T$. If it was, it would enter into direct contradiction with the definition of ${{BB}^+}_{{P'\textrm{´}}_T}$: the program ${P'\textrm{´}}_T$ will become undefined for an input, which by assumption is false. 

Further ahead the reader may attest that this function is the same as the one defined by $U\left({\pi' \textrm{´}}_{\mathrm{\Omega }}\circ {P'\textrm{´}}_T\circ 0^{\left|U\left(P'_{\sum }\circ {P'\textrm{´}}_T\circ N\right)\right|}1\circ U\left(P'_{\sum }\circ {P'\textrm{´}}_T\circ N\right)\right)$. This is deliberate, as it makes ${\pi \textrm{´}}_{\mathrm{\Omega }}$ and ${{BB}^+}_{{P'\textrm{´}}_T}\left(N\right)$ resemble ${\pi }_{\mathrm{\Omega }}$ and $BB\textrm{´}\left(N\right)$ respectively, as introduced in \cite{Abrahao2015}. This is what allows us to simulate some hypercomputable properties of a first order hypercomputer in relation to a computer.

\section[Psum]{Program $P_{\sum } $}

Let $P_{\sum}$ be a program that takes ${P'\textrm{´}}_T\circ N$ as input and calculates the binary sum of all algorithmic probabilities of the programs with size $\le N$ in the language $L$ that halt in computation time defined by ${P'\textrm{´}}_T$ (or, $U_{{P'\textrm{´}}_T}\left(w\right)\neq l_1$).  If $N=0,$ the output is $0$.

If ${P'\textrm{´}}_T\ $is not a total\textbf{\textit{ }}function (i.e., does not halt for an input $w$), then $P_{\sum }$ will also not be defined for all inputs. If ${P'\textrm{}}_T$ is, then obviously $P_{\sum }$ will also be.

Remember, this is a self-delimiting language; therefore, the output of $P_{\sum }$ will always be a real binary number between 0 and 1.

But what happens if we increase the number $N$ progressively? As the sum of the algorithmic probabilities of all $L$ programs is $\le 1$ and $\ U(P_{\sum }\circ {P'\textrm{´}}_T\circ N)\mathrm{\ }$is always non-decreasing, will there be a limit in $U(P_{\sum }\circ {P'\textrm{´}}_T\circ N)$ when $N$ tends to infinity (as long as ${P'\textrm{´}}_T$ is a total function). What does this limit reveal? It will be, precisely, the time-bounded version of Chaitin's Omega number.

\section[Relative Omega number]{A relative Chaitin's constant: the time-bounded halting probability}

Through the construction of program $P_{SM}$, it is possible to note that, in the limit, $P_{\sum }\ $calculates the sum of all programs in $L$ that halt in computation time determined by ${P'\textrm{´}}_T.\ \ $With this, a direct analogy with Chaitin's Omega number \cite{Calude2010, Chaitin1975, Calude2002} is obtained, which justifies calling this number a time-bounded halting probability.

Let ${P'\textrm{´}}_T\ $be a total\textbf{\textit{ }}function. So ${{\Omega}}_{{P'\textrm{´}}_T}$ may denote the \textbf{time-bounded} \textbf{halting probability} defined by:

\[{\mathrm{\Omega }}_{{P'\textrm{´}}_T}=\sum_{p\mathrm{\ is\ a\ program\ in\ }L\mathrm{\ that\ halts\ in\ compution\ time\ }\le U({P'\textrm{´}}_T\circ p).}{2^{-|p|}}\]

In other words, $p$ is counted in this sum if, and only if, for every $y$ and $x$,  

\[U\left(T\circ p\right)=x\ \wedge \ U\left({P'\textrm{´}}_T\circ p\right)=y\mathrm{\ }\to x\le y\]

Thus, the more rapidly increasing is the function given by program ${P'\textrm{´}}_T$, the closer $\mathrm{\Omega }$ will be approached. In addition, note that non-halting programs\textbf{\textit{ }}$p$\textbf{ }are excluded: after $y$ computations, if program \textit{p} has not halted yet, it can only either never halt, or halt in time $x$, with $x>y$.

\section[Subprograms for lower approximations]{Subprograms ${{\pi' }{\mathrm{\textrm{´}}}}_{{\mathrm{\Omega }}}{\mathrm{\circ }}{{P'}{\mathrm{\textrm{´}}}}_{{T}}{\mathrm{\circ }}{{\mathrm{0}}}^{\left|{\rho }\right|}{\mathrm{1}}{\mathrm{\circ }}{\rho }$}

Program ${\pi' \mathrm{\textrm{´}}}_{\mathrm{\Omega }}\mathrm{\circ }{P'\mathrm{\textrm{´}}}_T\mathrm{\circ }0^{\left|\rho \right|}\mathrm{1}\mathrm{\circ }\rho $ will now be explained. What does this program do? It is almost the same program\textit{ }${\pi }_{{\Omega}}{\ \ }$used by Chaitin in his early models for\textbf{\textit{ }}cumulative evolution in metabiology \cite{Chaitin2012}. However, we take here as reference a Turing submachine $U_{{P'\textrm{´}}_T}$ instead of the universal machine $U$. Furthermore, the output is not approximation indices to ${{\Omega}}_{{P'\textrm{´}}_T}$, but approximations to ${{BB}^+}_{{P'\textrm{´}}_T}$. This is equivalent if one is dealing with the classical halting probability ${\Omega}$ e the classical Busy Beaver function $BB(N)$, which are defined on $U$. However, it does not necessarily holds if one is dealing with Turing submachines.

Let ${P'\textrm{´}}_T$ be a total function. 

Program\textit{ }${\pi' \textrm{´}}_{{\Omega}}$\textbf{ }reads bit string ${P'\textrm{´}}_T\circ 0^{\left|\rho \right|}1\circ \rho $ as its input and sums all the algorithmic probabilities of programs of size $\le n$ that have halted in the computation time determined by ${P'\textrm{´}}_T$. It starts with $\ n=1\ $and continues up to $k$ so that the sum of all the algorithmic probabilities of all programs of size $\le k$ is equal to or larger than the value of $\rho .$ Then it calculates the largest output among\textbf{\textit{ }}these subprograms of size\textbf{\textit{ }}${\le }k$, adds 1 and returns this value. If $\rho =0$, then it returns $0$.

Note that if $\rho >{\mathrm{\Omega }}_{{P'\textrm{´}}_T}$, this program will never halt. In case $\rho ={\mathrm{\Omega }}_{{P'\textrm{´}}_T}$ --- which would be an ideal possibility --- ${\pi' \textrm{´}}_{\mathrm{\Omega }}\ $may, or may not, halt: it will only not ever halt if $\left|\rho \right|=|{\mathrm{\Omega }}_{{P'\textrm{´}}_T}|$ is infinite, since ${\pi' \textrm{´}}_{\mathrm{\Omega }}$ would never finish reading its input when ${\mathrm{\Omega }}_{{P'\textrm{´}}_T}\ $is of infinite size. If $\rho $ has finite size and $\rho \le {\mathrm{\Omega }}_{{P'\textrm{´}}_T},\ $then  ${\pi' \textrm{´}}_{\mathrm{\Omega }}$ halts.

Hence it may be concluded that ${\pi' \textrm{´}}_{\mathrm{\Omega }}$ takes a finite lower approximation to\textbf{\textit{ }}${\mathrm{\Omega }}_{{P'\textrm{´}}_T}$ and calculates a value that exceeds the value of any output produced, by any program involved, in ascending order of size, over the sum of algorithmic probabilities that results in the lower approximation in question. This is practically analogous to what happens with\textbf{\textit{ }}${\pi }_{\mathrm{\Omega }}\ $in relation to\textbf{\textit{ }}$\mathrm{\Omega }$ and $BB'\textrm{´}(N)$ in Chaitin's work \cite{Chaitin2013}, except for a increment (which is adding $1$).

Thus, assuming ${P'\textrm{´}}_T$ is total, it can be shown that ${\pi' \mathrm{\textrm{´}}}_{\mathrm{\Omega }}\mathrm{\circ }{P'\mathrm{\textrm{´}}}_T\mathrm{\circ }0^{\left|\rho' \mathrm{\textrm{´}}\right|}\mathrm{1}\mathrm{\circ }\rho' \mathrm{\textrm{´}}$ always halts when $\rho' \textrm{´}\le U\left(P_{\sum }\circ {P'\textrm{´}}_T\circ N\right)$, for every $N$.  And it does not matter if $|\rho' \textrm{´}|>|U\left(P_{\sum }\circ {P'\textrm{´}}_T\circ N\right)|$, as what is being dealt with are real numbers between 0 and 1. For example, the trailing zeros do not affect the fact that $\rho' \textrm{´}\le U\left(P_{\sum }\circ {P'\textrm{´}}_T\circ N\right)$ --- and this will be important later on.

\section[Non-diagonal Submachine]{Submachine $U_{{P^*}_T\circ {P'\textrm{´}}_T\circ P_T}$}

Now we will define a Turing submachine $U_{{P^*}_T\circ {P'\textrm{´}}_T\circ P_T}$. Note that ${P'\textrm{´}}_T$ and $P_T$  are two total functions, by assumption.  This will remain until we reach submachine $U_{{P^{**}}_T\circ P_T}$, where $P_T$ is still total, by assumption, but in place of ${P'\textrm{´}}_T$, it will be self-referentially put ${P^{**}}_T\circ P_T$.  Also remember that program $T$ receives $w$ as input and returns the computation time of $U\left(w\right)$.

Program ${P^*}_T\ $receives ${P'\textrm{´}}_T$, $P_T$ and $w$ as inputs, and returns:

\begin{enumerate}[(i)]
\item  $U\mathrm{(}T\mathrm{\circ }w\mathrm{)}$, if $w\ $is in the form$\mathrm{\ }{\pi' \mathrm{\textrm{´}}}_{\mathrm{\Omega }}\mathrm{\circ }{P'\mathrm{\textrm{´}}}_T\mathrm{\circ }0^{\left|\rho \right|}\mathrm{1}\mathrm{\circ }\rho$ and

\[U\left(P_{\mathrm{\sum }}\mathrm{\circ }{P'\mathrm{\textrm{´}}}_T\mathrm{\circ }\mathrm{(|}{\pi' \mathrm{\textrm{´}}}_{\mathrm{\Omega }}\mathrm{\circ }{P'\mathrm{\textrm{´}}}_T\mathrm{\circ }0^{\left|\rho \right|}\mathrm{1}\mathrm{\circ }\rho \mathrm{|-1)}\right)\mathrm{\ge }\rho \]

and

\[U\left(T\mathrm{\circ }w\right)\mathrm{>}U\left(P_T\mathrm{\circ }w\right)\mathrm{;}\]

\item   $U(P_T\circ w)$, if case (i) does not apply; \textbf{\textit{ }}
\end{enumerate}

Here a program is being defined upon two total functions: ${P'\textrm{´}}_T$ and $P_T$. The first step to verify whether or not ${P^*}_T$  is total is noting that there is a program that always decides if $w$ lies in case (i) or (ii). 

The program must first verify the form of the program $w$. For case (i), if 

\[U\left(P_{\sum }\circ {P'\textrm{´}}_T\circ (|{\pi' \textrm{´}}_{\mathrm{\Omega }}\circ {P'\textrm{´}}_T\circ 0^{\left|\rho \right|}1\circ \rho |-1)\right)\ge \rho\]

\noindent we will have that $U\left(T\circ {\pi' \textrm{´}}_{\mathrm{\Omega }}\circ {P'\textrm{´}}_T\circ 0^{\left|\rho \right|}1\circ \rho \mathrm{\ }\right)$ is always well defined. The only thing left to determine is whether

\[U\left(T\circ {\pi' \textrm{´}}_{\mathrm{\Omega }}\circ {P'\textrm{´}}_T\circ 0^{\left|\rho \right|}1\circ \rho \mathrm{\ }\right)>U\left(P_T\circ {\pi' \textrm{´}}_{\mathrm{\Omega }}\circ {P'\textrm{´}}_T\circ 0^{\left|\rho \right|}1\circ \rho \mathrm{\ }\right)\]

\noindent or not, which is also decidable. On the other hand, if 

\[U\left(P_{\sum }\circ {P'\textrm{´}}_T\circ (|{\pi' \textrm{´}}_{\mathrm{\Omega }}\circ {P'\textrm{´}}_T\circ 0^{\left|\rho \right|}1\circ \rho |-1)\right)<\rho\]

\noindent then case (ii) applies directly. Case (ii) derives directly from the fact that $P_T$ is also total.

It is important to attend to the fact that ${P^*}_T\circ {P'\textrm{´}}_T\circ P_T$ is sufficient computation time to compute anything that $U_{P_T}$ may compute. In case (ii) this is obvious. In case of an (i)-formed program, condition $U\left(T\circ w\right)>U(P_T\circ w)$ will always be present ensuring that ${P^*}_T\circ {P'\textrm{´}}_T\circ P_T$ is always an extension of $P_T$, and never a restriction.

Summarizing what ${P^*}_T\circ {P'\textrm{´}}_T\circ P_T\ $does in case (i), $U_{{P^*}_T\circ {P'\textrm{´}}_T\circ P_T}$ always allows, at least, a program of the form ${\pi' \textrm{´}}_{\mathrm{\Omega }}\circ {P'\textrm{´}}_T\circ 0^{\left|\rho \right|}1\circ \rho $, which will calculate a value always higher than any output from any program on $U_{{P'\textrm{´}}_T}$ of size $\le N$, so that $U\left(P_{\sum }\circ {P'\textrm{´}}_T\circ (|{\pi' \textrm{´}}_{\mathrm{\Omega }}\circ {P'\textrm{´}}_T\circ 0^{\left|\rho \right|}1\circ \rho |-1)\right)\ge U\left(P_{\sum }\circ {P'\textrm{´}}_T\circ (N)\right)$. As we will prove, this essentially always enables submachine $U_{{P^*}_T\circ {P'\textrm{´}}_T\circ P_T}$ to calculate ${{BB}^+}_{{P'\textrm{´}}_T}\left(N\right)$ with a subprogram of size $\ge N+1$ and $\le 2N+C$,  where $C$ is a constant\textbf{\textit{.}}

Of course, once that for every $w$, $U\left(P_T\mathrm{\circ }w\right)\mathrm{\le }U\mathrm{(}{P^{\mathrm{*}}}_T\mathrm{\circ }{P'\mathrm{\textrm{´}}}_T\mathrm{\circ }P_T\mathrm{\circ }w\mathrm{)}$, then

 \[{\mathrm{\Omega }}_{{P^{\mathrm{*}}}_T\mathrm{\circ }{P'\mathrm{\textrm{´}}}_T\mathrm{\circ }P_T}\mathrm{\ge }{\mathrm{\Omega }}_{P_T}\]
 
 \noindent However, what about if our objective is for $U\left(P_{\mathrm{\sum }}\mathrm{\circ }{P'\mathrm{\textrm{´}}}_T\mathrm{\circ }N\right)$ to always be an approximation to ${\mathrm{\Omega }}_{{P^{\mathrm{*}}}_T\mathrm{\circ }{P'\mathrm{\textrm{´}}}_T\mathrm{\circ }P_T}$ itself$\mathrm{,\ }$rather than an approximation to ${\mathrm{\Omega }}_{{P'\mathrm{\textrm{´}}}_T}$? That is, in case (i) that $U_{{P^{\mathrm{*}}}_T\mathrm{\circ }{P'\mathrm{\textrm{´}}}_T\mathrm{\circ }P_T}$ always be able to calculate values of ${{BB}^{\mathrm{+}}}_{{P^{\mathrm{*}}}_T\mathrm{\circ }{P'\mathrm{\textrm{´}}}_T\mathrm{\circ }P_T}$ with a subprogram of size $\mathrm{\ge }N\mathrm{+1}$? Which program should replace for ${P'\mathrm{\textrm{´}}}_T$? Here the employment of a self-diagonalizing self-reference will become evident, within program ${P^{\mathrm{*}}}_T$ itself, to build program ${P^{\mathrm{**}}}_T$  below.

\section[Diagonal Submachine]{Submachine $U_{{{P^{**}}_T}\circ P_T}$}

Let $P_T$ be a total function. Program ${P^{\mathrm{**}}}_T$ receives $P_T$ and $w$ as inputs, reads itself, $P_T$ and $w$, and then  assembles  ${P^{\mathrm{*}}}_T\mathrm{\circ }{P^{\mathrm{**}}}_T\mathrm{\circ }P_T\mathrm{\circ }P_T\mathrm{\circ }w$ and returns $U\mathrm{(}{P^{\mathrm{*}}}_T\mathrm{\circ }{P^{\mathrm{**}}}_T\mathrm{\circ }P_T\mathrm{\circ }P_T\mathrm{\circ }w\mathrm{)}$.

That is, $U\left({P^{\mathrm{**}}}_T\mathrm{\circ }P_T\mathrm{\circ }w\right)\mathrm{=}U\mathrm{(}{P^{\mathrm{*}}}_T\mathrm{\circ }{P^{\mathrm{**}}}_T\mathrm{\circ }P_T\mathrm{\circ }P_T\mathrm{\circ }w\mathrm{)}$, returning:

\begin{enumerate}[(i)]
	\item  $U\mathrm{(}T\mathrm{\circ }w\mathrm{)}$, if $w$ is in form ${\pi \mathrm{\textrm{´}}}_{\mathrm{\Omega }}\mathrm{\circ }{P^{\mathrm{**}}}_T\mathrm{\circ }P_T\mathrm{\circ }0^{\left|\rho \right|}\mathrm{1}\mathrm{\circ }\rho $ and

\[U\left(P_{\mathrm{\sum }}\mathrm{\circ }{P^{\mathrm{**}}}_T\mathrm{\circ }P_T\mathrm{\circ }\mathrm{(|}{\pi \mathrm{\textrm{´}}}_{\mathrm{\Omega }}\mathrm{\circ }{P^{\mathrm{**}}}_T\mathrm{\circ }P_T\mathrm{\circ }0^{\left|\rho \right|}\mathrm{1}\mathrm{\circ }\rho \mathrm{|-1)}\right)\mathrm{\ge }\rho \] 

and

\[U\left(T\mathrm{\circ }w\right)\mathrm{>}U\left(P_T\mathrm{\circ }w\right)\mathrm{;}\]

\item  $U(P_T\circ w)$, if case (i) does not apply;

\end{enumerate}

The issue at hand is whether this program is a total function or not. It will not be as simple as in the previous case. Case (ii) remains trivial, since it depends only on $P_T.\ $In (i), the problem is that when a self-reference is introduced, a program might fall into an endless loop. To prove that this self-reference in ${P^{**}}_T$ will never generate an endless loop, for any $w$, a mathematical induction proof\textbf{\textit{ }}will be built. 

The simple ``trick'' to this demonstration is realizing that for any program of size $N$ on $U_{{P^{**}}_T\circ P_{T\ \ }}$to be well defined, it will merely depend on the programs in $U_{{P^{**}}_T\circ P_T}$ of size $<N$, or on  $P_T$. Note that, as ${\pi \textrm{´}}_{\mathrm{\Omega }}$ is predetermined, it has a minimum size, thus there will be a maximum size $M_{(i)}$ for $w$ such that if $\left|w\right|\le M_{(i)}$, then $U\left({P^{**}}_T\circ P_T\circ w\right)\ $will never fall under case (i). With this, since $P_T$ is given as total\textbf{\textit{, }}then ${P^{**}}_T\circ P_T$ will always be well defined for sufficiently small $w$. To conclude the proof by induction, it is sufficient to note that if ${P^{**}}_T\circ P_T$ is well defined for every $w$ of size $\le k$, so ${P^{**}}_T\circ P_T$ will be well defined for every $w$ of size $k+1$. These are the key ideas of what will be done in theorem \ref{thm10.1} below.

Once ${P^{**}}_T\circ P_T$ is total,  ${{BB}^+}_{{P^{**}}_T\circ P_T}(N)$ and ${{\Omega}}_{{P^{**}}_T\circ P_T}\ $can be defined, and therefore, as an immediate corollary of theorem \ref{thm11.1}, ${{BB}^+}_{{P^{**}}_T\circ P_T}(N)$ will be as uncomputable and as incompressible in relation to $U_{{P^{**}}_T\circ P_T}$ as $BB\textrm{´}(N)$ is in relation to $U$ --- except for\textbf{\textit{ }}a constant. Note, thus, that ${{BB}^+}_{{P^{**}}_T\circ P_T}(N)$ also serve to ``measure'' the algorithmic complexity of programs on $U_{{P^{**}}_T\circ P_T}$: no output equal to or higher than ${{BB}^+}_{{P^{**}}_T\circ P_T}(N)$ can be achieved by any  subprogram of size $\le N$, but can always be calculated by a subprogram of size $\le 2N+C$. See theorem \ref{thm10.2}.

The essential hypercomputational property that will thus be able to be simulated is that there are programs in $U_{{P^{**}}_T\circ P_T}$ that use finite lower\textbf{\textit{ }}approximations to ${{\Omega}}_{{P^{**}}_T\circ P_T}$ to occasionally calculate ${{BB}^+}_{{P^{**}}_T\circ P_T}$ values. At least regarding this aspect, it is as if the machine that computes function ${{BB}^+}_{{P^{**}}_T\circ P_T}(N)$ was a first order oracle Turing machine, only that now in relation to submachine $U_{{P^{**}}_T\circ P_T}$ and not in relation to $U.$

\newtheorem{thm}{Theorem}[section] 
\newproof{pf}{Proof}
\begin{thm}\label{thm10.1} 
	
	\textrm{\newline}
	Let $P_T$\textit{ }be a total function.
	Then, for every $w\in L$, $U({P^{**}}_T\circ P_T\circ w)$ is a well-defined value, i.e. $U({P^{**}}_T\circ P_T\circ w)$ is a total function of $w;$ or $U_{{P^{**}}_T\circ P_T}$ is a Turing submachine.

	\begin{pf}
		\noindent
			
		\begin{enumerate}[a)] 
		
		\item  Case valid for every $w$ where $|w|\le k_0$:\newline 
			
		As ${\pi \textrm{´}}_{\mathrm{\Omega }}$ is a known program, it has a size (in the language $L$ of our choice).
		
		We also know program ${P^{**}}_T\circ P_T$, which will also have, therefore, a size.
		
		Hence,
		
		\[\mathrm{\exists }k_0\mathrm{(\ }k_0\mathrm{=}{\mathrm{min} \left\{\left|{\pi' \mathrm{\textrm{´}}}_{\mathrm{\Omega }}\mathrm{\circ }{P^{\mathrm{**}}}_T\mathrm{\circ }P_T\mathrm{\circ }w\right|\mathrm{;}\mathrm{\ \ }w\mathrm{\in }L\right\}\ }\mathrm{-}\mathrm{1)}\] 
		
		 and
		
		\[\mathrm{\forall }w\mathrm{(}\left|w\right|\mathrm{\le }\mathrm{\ }k_0\boldsymbol {\mathrm{\to }} w\mathrm{\neq }{\pi' \mathrm{\textrm{´}}}_{\mathrm{\Omega }}\mathrm{\circ }{P^{\mathrm{**}}}_T\mathrm{\circ }P_T\mathrm{\circ }0^{\left|\rho \right|}\mathrm{1}\mathrm{\circ }\rho \mathrm{\ }\] 
		
		Through the definition of language $L$, determining whether 
		
		\[w={\pi' \textrm{´}}_{\mathrm{\Omega }}\circ {P^{**}}_T\circ P_T\circ 0^{\left|\rho \right|}1\circ \rho \ \vee w\neq {\pi' \textrm{´}}_{\mathrm{\Omega }}\circ {P^{**}}_T\circ P_T\circ 0^{\left|\rho \right|}1\circ \rho \]
		
		is a decidable problem. Thus, ${P^{**}}_T$ can always recursively determine which option applies.
		
		Thus,
		
		\[\mathrm{\forall }w\mathrm{(}\left|w\right|\mathrm{\le }\mathrm{\ }k_0\boldsymbol{\mathrm{\to }}U\left({P^{\mathrm{**}}}_T\mathrm{\circ }P_T\mathrm{\circ }w\right)\mathrm{=}U\mathrm{(}P_T\mathrm{\circ }w\mathrm{)}\mathrm{)}\]

		Since, by our initial assumption, $P_T$ is a total function, then
		
		\[\mathrm{\forall }w\mathrm{\exists }y\mathrm{(}\left|w\right|\mathrm{\le }\mathrm{\ }k_0\boldsymbol{\mathrm{\to }}U\left({P^{\mathrm{**}}}_T\mathrm{\circ }P_T\mathrm{\circ }w\right)\mathrm{=}y\mathrm{)}\] 
		
		This concludes the first part of the mathematical induction.\newline

		\item Case valid for every $w$ where $\ |w|\le k$:\newline 
		
		Our objective is to demonstrate that, in this case, it will also be valid for every $w$ where $\left|w\right|=k+1$. 
		
		First, take any arbitrary program $w$ of size $k+1$.
		
		We have that either $w$ is in form ${\pi' \textrm{´}}_{\mathrm{\Omega }}\circ {P^{**}}_T\circ P_T\circ 0^{\left|\rho \right|}1\circ \rho $ or in any other form. 
		
		As in case a), knowing whether 
		
		\[w={\pi' \textrm{´}}_{\mathrm{\Omega }}\circ {P^{**}}_T\circ P_T\circ 0^{\left|\rho \right|}1\circ \rho \ \vee w\neq {\pi' \textrm{´}}_{\mathrm{\Omega }}\circ {P^{**}}_T\circ P_T\circ 0^{\left|\rho \right|}1\circ \rho\] 
		
		is a decidable problem. Thus, ${P^{**}}_T$ can always determine correctly any of these options. Below each possibility will be examined separately.\newline
		
		\begin{enumerate}[(i)]
		
			\item  In case $w\mathrm{=}{\pi' \mathrm{\textrm{´}}}_{\mathrm{\Omega }}\mathrm{\circ }{P^{\mathrm{**}}}_T\mathrm{\circ }P_T\mathrm{\circ }0^{\left|\rho \right|}\mathrm{1}\mathrm{\circ }\rho $:\newline
		
		 It will suffice for ${P^{\mathrm{**}}}_T\ $ to calculate 
		 
		 \[\mathrm{\ }U\left(P_{\mathrm{\sum }}\mathrm{\circ }{P^{\mathrm{**}}}_T\mathrm{\circ }P_T\mathrm{\circ }\mathrm{(|}{\pi' \mathrm{\textrm{´}}}_{\mathrm{\Omega }}\mathrm{\circ }{P^{\mathrm{**}}}_T\mathrm{\circ }P_T\mathrm{\circ }0^{\left|\rho \right|}\mathrm{1}\mathrm{\circ }\rho \mathrm{|-1)}\right)\] 
		 
		 and verify whether 
		 
		 \[U\left(P_{\mathrm{\sum }}\mathrm{\circ }{P^{\mathrm{**}}}_T\mathrm{\circ }P_T\mathrm{\circ }\mathrm{(|}{\pi' \mathrm{\textrm{´}}}_{\mathrm{\Omega }}\mathrm{\circ }{P^{\mathrm{**}}}_T\mathrm{\circ }P_T\mathrm{\circ }0^{\left|\rho \right|}\mathrm{1}\mathrm{\circ }\rho \mathrm{|-1)}\right)\mathrm{\ge }\rho\]
		 
		 The crux of the matter is to prove that $U\left({P^{\mathrm{**}}}_T\mathrm{\circ }P_T\mathrm{\circ }w\right)$ needs to be defined for every $w$,  wherein $\left|w\right|\mathrm{\le }\left|{\pi' \mathrm{\textrm{´}}}_{\mathrm{\Omega }}\mathrm{\circ }{P^{\mathrm{**}}}_T\mathrm{\circ }P_T\mathrm{\circ }0^{\left|\rho \right|}\mathrm{1}\mathrm{\circ }\rho \right|\mathrm{-}\mathrm{1}$, so that $P_{\mathrm{\sum }}\mathrm{\circ }{P^{\mathrm{**}}}_T\mathrm{\circ }P_T\mathrm{\circ }\mathrm{(|}{\pi' \mathrm{\textrm{´}}}_{\mathrm{\Omega }}\mathrm{\circ }{P^{\mathrm{**}}}_T\mathrm{\circ }P_T\mathrm{\circ }0^{\left|\rho \right|}\mathrm{1}\mathrm{\circ }\rho \mathrm{|-1)}$  always halts\textbf{\textit{.}}
		 By the inductive hypothesis, $U\left({P^{\mathrm{**}}}_T\mathrm{\circ }P_T\mathrm{\circ }w\right)$ is well defined for every $w$, with $\left|w\right|\mathrm{\le }k$. Therefore, since 
		 
		 \[\left|{\pi' \mathrm{\textrm{´}}}_{\mathrm{\Omega }}\mathrm{\circ }{P^{\mathrm{**}}}_T\mathrm{\circ }P_T\mathrm{\circ }0^{\left|\rho \right|}\mathrm{1}\mathrm{\circ }\rho \right|\mathrm{=}k\mathrm{+1}\]
		 
		 then $U\left({P^{\mathrm{**}}}_T\mathrm{\circ }P_T\mathrm{\circ }w\right)$ is defined for every $w$ where 
		 
		 \[\left|w\right|\mathrm{\le }\left|{\pi' \mathrm{\textrm{´}}}_{\mathrm{\Omega }}\mathrm{\circ }{P^{\mathrm{**}}}_T\mathrm{\circ }P_T\mathrm{\circ }0^{\left|\rho \right|}\mathrm{1}\mathrm{\circ }\rho \right|\mathrm{-}\mathrm{1=}k\]
		 
		 Thus\textbf{\textit{,}} when ${P^{\mathrm{**}}}_T$ runs $P_{\mathrm{\sum }}\mathrm{\circ }{P^{\mathrm{**}}}_T\mathrm{\circ }P_T\mathrm{\circ }\mathrm{(|}{\pi' \mathrm{\textrm{´}}}_{\mathrm{\Omega }}\mathrm{\circ }{P^{\mathrm{**}}}_T\mathrm{\circ }P_T\mathrm{\circ }0^{\left|\rho \right|}\mathrm{1}\mathrm{\circ }\rho \mathrm{|-1)}$, it will halt, never entering into a loop.\textbf{\textit{}}
		 Moreover, as $\rho $ is given by $w$, determining whether or not 
		 
		 \[U\left(P_{\mathrm{\sum }}\mathrm{\circ }{P^{\mathrm{**}}}_T\mathrm{\circ }P_T\mathrm{\circ }\mathrm{(|}{\pi' \mathrm{\textrm{´}}}_{\mathrm{\Omega }}\mathrm{\circ }{P^{\mathrm{**}}}_T\mathrm{\circ }P_T\mathrm{\circ }0^{\left|\rho \right|}\mathrm{1}\mathrm{\circ }\rho \mathrm{|-1)}\right)\mathrm{\ge }\rho  \]
		 
		 will be a decidable problem. Thus, ${P^{**}}_T$ can recursively verify whether
		  
		 \[U\left(P_{\sum }\circ {P^{**}}_T\circ P_T\circ (|{\pi' \textrm{´}}_{\mathrm{\Omega }}\circ {P^{**}}_T\circ P_T\circ 0^{\left|\rho \right|}1\circ \rho |-1)\right)\ge \rho \]
		 
		 In case $\left(P_{\sum }\circ {P^{**}}_T\circ P_T\circ (|{\pi' \textrm{´}}_{\mathrm{\Omega }}\circ {P^{**}}_T\circ P_T\circ 0^{\left|\rho \right|}1\circ \rho |-1)\right)\ge \rho $, by the definition of ${\pi' \mathrm{\textrm{´}}}_{\mathrm{\Omega }}$, we have that for every $\rho' \mathrm{\textrm{´}}$, if 
		 
		 \[U\left(P_{\mathrm{\sum }}\mathrm{\circ }{P^{\mathrm{**}}}_T\mathrm{\circ }P_T\mathrm{\circ }\mathrm{(|}{\pi' \mathrm{\textrm{´}}}_{\mathrm{\Omega }}\mathrm{\circ }{P^{\mathrm{**}}}_T\mathrm{\circ }P_T\mathrm{\circ }0^{\left|\rho \right|}\mathrm{1}\mathrm{\circ }\rho' \mathrm{\textrm{´}|-1)}\right)\mathrm{\ge }\rho' \mathrm{\textrm{´}}\]
		 
		 then  ${\pi' \mathrm{\textrm{´}}}_{\mathrm{\Omega }}\mathrm{\circ }{P^{\mathrm{**}}}_T\mathrm{\circ }P_T\mathrm{\circ }0^{\left|\rho \right|}\mathrm{1}\mathrm{\circ }\rho' \mathrm{\textrm{´}}$ will always halt. \noindent
		 Thus\textbf{\textit{,}} if
		  
		 \[U\left(P_{\mathrm{\sum }}\mathrm{\circ }{P^{\mathrm{**}}}_T\mathrm{\circ }P_T\mathrm{\circ }\mathrm{(|}{\pi' \mathrm{\textrm{´}}}_{\mathrm{\Omega }}\mathrm{\circ }{P^{\mathrm{**}}}_T\mathrm{\circ }P_T\mathrm{\circ }0^{\left|\rho \right|}\mathrm{1}\mathrm{\circ }\rho \mathrm{|-1)}\right)\mathrm{\ge }\rho\]
		 
		 then $U\left(T\mathrm{\circ }w\right)\mathrm{=}U\left(T\mathrm{\circ }{\pi' \mathrm{\textrm{´}}}_{\mathrm{\Omega }}\mathrm{\circ }{P^{\mathrm{**}}}_T\mathrm{\circ }P_T\mathrm{\circ }0^{\left|\rho \right|}\mathrm{1}\mathrm{\circ }\rho \right)$ will always be well defined.
		
		 In case $\ \left(P_{\sum }\circ {P^{**}}_T\circ P_T\circ (|{\pi' \textrm{´}}_{\mathrm{\Omega }}\circ {P^{**}}_T\circ P_T\circ 0^{\left|\rho \right|}1\circ \rho |-1)\right)<\rho $, we will proceed to case (ii).
		
		 Furthermore, due to the fact that $P_T$ is total, one can always decide whether or not 
		 
		 \[U\left(T\circ w\right)>U(P_T\circ w) \]
		
			Thus,\textbf{\textit{ }} $U_{{P^{\mathrm{**}}}_T\mathrm{\circ }P_T}\left(\mathrm{w}\right)$ \textbf{\textit{ }}will always be well defined if $w$ falls into case (i).\textbf{\textit{}}
		
		\noindent \textbf{\textit{}}
		
		\noindent \textbf{\textit{}}
		
			\item If case (i) does not apply to $w$:\newline

		It follows directly from the fact that $P_T\ $is total.\textbf{\textit{}}
		
		Thus,\textbf{\textit{ }}from (i) and (ii), for every $w$ where $\left|w\right|=k+1$, $U\left({P^{**}}_T\circ P_T\circ w\right)$ will be well defined.\newline
		
		\end{enumerate}

	\item  Finishing the induction:\newline
			
	Through\textbf{\textit{ }}a) and b) it is concluded, by induction, that $U\left({P^{**}}_T\circ P_T\circ w\right)\ $is well defined for every $w\in L$.\newline

	\end{enumerate}
		
	\end{pf}
	
\end{thm}

\newproof{kipf}{The key idea of the proof}
\begin{thm} \label{thm10.2}	
\textrm{\newline}	
Let $N$ be a natural number. Let $P_T$ be a total function. Then, for\textbf{\textit{ }}every $N$, there is a program in the form  

\[{\pi' \mathrm{\textrm{´}}}_{\mathrm{\Omega }}\mathrm{\circ }{P^{\mathrm{**}}}_T\mathrm{\circ }P_T\mathrm{\circ }0^{\left|\rho \right|}\mathrm{1}\mathrm{\circ }\rho \]

\noindent such that
	
\[U_{{P^{\mathrm{**}}}_T\mathrm{\circ }P_T}\left({\pi' \mathrm{\textrm{´}}}_{\mathrm{\Omega }}\mathrm{\circ }{P^{\mathrm{**}}}_T\mathrm{\circ }P_T\mathrm{\circ }0^{\left|\rho \right|}\mathrm{1}\mathrm{\circ }\rho \right)\mathrm{\ge }{{BB}^{\mathrm{+}}}_{{P^{\mathrm{**}}}_T\mathrm{\circ }P_T}\left(N\right)\] 
		
\noindent and 
	
\[\left|{\pi' \mathrm{\textrm{´}}}_{\mathrm{\Omega }}\mathrm{\circ }{P^{\mathrm{**}}}_T\mathrm{\circ }P_T\mathrm{\circ }0^{\left|\rho \right|}\mathrm{1}\mathrm{\circ }\rho \right|\le 2N+C\] 
	
\noindent where $C$ is a constant.\newline

\begin{kipf}
	
\textnormal{ \newline 
	\newline 
	\indent Let us use the fact that ${\mathrm{\Omega }}_{{P^{**}}_T\circ P_T}$ is a number that contains the necessary information on programs on $U_{{P^{**}}_T\circ P_T}\ $in order to compute function ${{BB}^+}_{{P^{**}}_T\circ P_T}\left(N\right)$. This is purposefully analogous to $\mathrm{\Omega }$ and $BB(N)$. Furthermore, ${P^{**}}_T\ $ was constructed to allow lower approximations to  ${\mathrm{\Omega }}_{{P^{**}}_T\circ P_T}$ to be used for $U_{{P^{**}}_T\circ P_T}$ to calculate ${{BB}^+}_{{P^{**}}_T\circ P_T}\left(N\right)$.} \newline

\end{kipf}

\begin{pf}
\textnormal{\newline}

Through the previous theorem \ref{thm10.1}, it was proven that for any input $w$, ${P^{**}}_T\circ P_T\circ w$ halts. That is, $\forall w\exists y(\ U\left({P^{**}}_T\circ P_T\circ w\right)=y\ )$. It follows then that function $U(P_{\sum{}}\circ {P^{**}}_T\circ P_T\circ N)$ will also be well defined for every $N$.

Let then $N$ be an arbitrary natural number.

We will have that $U(P_{\sum{}}\circ {P^{**}}_T\circ P_T\circ N)$ give us a real binary number between 0 and 1, which is a finite lower\textbf{\textit{ }}approximation to ${\mathrm{\Omega }}_{{P^{**}}_T\circ P_T}$.

Thus, from the definition of program ${\pi' \textrm{´}}_{\mathrm{\Omega }}$,
 
\[U\left({\pi' \mathrm{\textrm{´}}}_{\mathrm{\Omega }}\mathrm{\circ }{P^{\mathrm{**}}}_T\mathrm{\circ }P_T\mathrm{\circ }0^{\left|U\left(P_{\mathrm{\sum }}\mathrm{\circ }{P^{\mathrm{**}}}_T\mathrm{\circ }P_T\mathrm{\circ }N\right)\right|}\mathrm{1}\mathrm{\circ }U\left(P_{\mathrm{\sum }}\mathrm{\circ }{P^{\mathrm{**}}}_T\mathrm{\circ }P_T\mathrm{\circ }N\right)\right)\mathrm{=}{{BB}^{\mathrm{+}}}_{{P^{\mathrm{**}}}_T\mathrm{\circ }P_T}\mathrm{(}N\mathrm{)}\] 

\noindent Also, clause (i) in the definition of program ${P^{**}}_T$ assures us that if 

\[U\left(P_{\mathrm{\sum }}\mathrm{\circ }{P^{\mathrm{**}}}_T\mathrm{\circ }P_T\mathrm{\circ }\mathrm{(|}{\pi' \mathrm{\textrm{´}}}_{\mathrm{\Omega }}\mathrm{\circ }{P^{\mathrm{**}}}_T\mathrm{\circ }P_T\mathrm{\circ }0^{\left|\rho \right|}\mathrm{1}\mathrm{\circ }\rho \mathrm{|-1)}\right)\mathrm{\ge }\rho \] 

\noindent then

\[U_{{P^{\mathrm{**}}}_T\mathrm{\circ }P_T}\left({\pi' \mathrm{\textrm{´}}}_{\mathrm{\Omega }}\mathrm{\circ }{P^{\mathrm{**}}}_T\mathrm{\circ }P_T\mathrm{\circ }0^{\left|\rho \right|}\mathrm{1}\mathrm{\circ }\rho \right)\mathrm{=}U\mathrm{(}{\pi' \mathrm{\textrm{´}}}_{\mathrm{\Omega }}\mathrm{\circ }{P^{\mathrm{**}}}_T\mathrm{\circ }P_T\mathrm{\circ }0^{\left|\rho \right|}\mathrm{1}\mathrm{\circ }\rho \mathrm{)}\]

\noindent That is why we need to make $\mathrm{|}{\pi' \mathrm{\textrm{´}}}_{\mathrm{\Omega }}\mathrm{\circ }{P^{\mathrm{**}}}_T\mathrm{\circ }P_T\mathrm{\circ }0^{\left|\rho \right|}\mathrm{1}\mathrm{\circ }\rho \mathrm{|-1}$  greater than or equal to $N$.

Let us, then, increase the size of $U(P_{\sum{}}\circ {P^{**}}_T\circ P_T\circ N)$, if necessary, through the following procedure: take the real number between 0 and 1 given by $U(P_{\sum{}}\circ {P^{**}}_T\circ P_T\circ N)$ and add (if necessary) enough trailing zeros until

\[\left|{\pi' \mathrm{\textrm{´}}}_{\mathrm{\Omega }}\mathrm{\circ }{P^{\mathrm{**}}}_T\mathrm{\circ }P_T\mathrm{\circ }0^{\left|U\left(P_{\sum{}}\mathrm{\circ }{P^{\mathrm{**}}}_T\mathrm{\circ }P_T\mathrm{\circ }N\right)\mathrm{*}\mathrm{0\dots 0}\right|}\mathrm{1}\mathrm{\circ }U\left(P_{\sum{}}\mathrm{\circ }{P^{\mathrm{**}}}_T\mathrm{\circ }P_T\mathrm{\circ }N\right)\mathrm{*}\mathrm{0\dots 0}\right|\mathrm{-}\mathrm{1}\mathrm{\ge }N\] 

\noindent We also have that, for every $k\mathrm{,\ }k'\mathrm{\textrm{´},\ }$ if 

\[k'\mathrm{\textrm{´}}\mathrm{\ge }k\]

\noindent then 

\[U\left(P_{\sum{}}\mathrm{\circ }{P^{\mathrm{**}}}_T\mathrm{\circ }P_T\mathrm{\circ }k'\mathrm{\textrm{´}}\right)\mathrm{\ge }U\left(P_{\sum{}}\mathrm{\circ }{P^{\mathrm{**}}}_T\mathrm{\circ }P_T\mathrm{\circ }k\right)\] 

\noindent Therefore,

\begin{displaymath}
\begin{split}
&U\biggl( P_{\mathrm{\sum }}\mathrm{\circ }{P^{\mathrm{**}}}_T\mathrm{\circ }P_T\mathrm{\circ }\mathrm{(|}{\pi' \mathrm{\textrm{´}}}_{\mathrm{\Omega }}\mathrm{\circ }{P^{\mathrm{**}}}_T\mathrm{\circ }P_T\mathrm{\circ }0^{\left|U\left(P_{\sum{}}\mathrm{\circ }{P^{\mathrm{**}}}_T\mathrm{\circ }P_T\mathrm{\circ }N\right)\mathrm{*}\mathrm{0\dots 0}\right|}\mathrm{1}\mathrm{\circ} \\ 
&\quad \;\,\circ U\left(P_{\sum{}}\mathrm{\circ }{P^{\mathrm{**}}}_T\mathrm{\circ }P_T\mathrm{\circ }N\right)\mathrm{*}\mathrm{0\dots 0|-1)}\biggr) \mathrm{\ge }\\
&U\left(P_{\sum{}}\mathrm{\circ }{P^{\mathrm{**}}}_T\mathrm{\circ }P_T\mathrm{\circ }N\right)\mathrm{=}\\
&U\left(P_{\sum{}}\mathrm{\circ }{P^{\mathrm{**}}}_T\mathrm{\circ }P_T\mathrm{\circ }N\right)\mathrm{*}\mathrm{0\dots 0}
\end{split}
\end{displaymath}

Thus, by the clause (i), as we have already remarked,

\begin{displaymath}
\begin{split}
&U_{{P^{\mathrm{**}}}_T\mathrm{\circ }P_T}\left({\pi' \mathrm{\textrm{´}}}_{\mathrm{\Omega }}\mathrm{\circ }{P^{\mathrm{**}}}_T\mathrm{\circ }P_T\mathrm{\circ }0^{\left|U\left(P_{\sum{}}\mathrm{\circ }{P^{\mathrm{**}}}_T\mathrm{\circ }P_T\mathrm{\circ }N\right)\mathrm{*}\mathrm{0\dots 0}\right|}\mathrm{1}\mathrm{\circ }U\left(P_{\sum{}}\mathrm{\circ }{P^{\mathrm{**}}}_T\mathrm{\circ }P_T\mathrm{\circ }N\right)\mathrm{*}\mathrm{0\dots 0}\right)\mathrm{=}\\
&U\mathrm{(}{\pi' \mathrm{\textrm{´}}}_{\mathrm{\Omega }}\mathrm{\circ }{P^{\mathrm{**}}}_T\mathrm{\circ }P_T\mathrm{\circ }0^{\left|U\left(P_{\sum{}}\mathrm{\circ }{P^{\mathrm{**}}}_T\mathrm{\circ }P_T\mathrm{\circ }N\right)\mathrm{*}\mathrm{0\dots 0}\right|}\mathrm{1}\mathrm{\circ }U\left(P_{\sum{}}\mathrm{\circ }{P^{\mathrm{**}}}_T\mathrm{\circ }P_T\mathrm{\circ }N\right)\mathrm{*}\mathrm{0\dots 0}\mathrm{)}
\end{split}
\end{displaymath} 

\noindent But, since $U\left(P_{\sum{}}\mathrm{\circ }{P^{\mathrm{**}}}_T\mathrm{\circ }P_T\mathrm{\circ }N\right)\mathrm{*}\mathrm{0\dots 0}$ and $U\left(P_{\sum{}}\mathrm{\circ }{P^{\mathrm{**}}}_T\mathrm{\circ }P_T\mathrm{\circ }N\right)$ are two equal real numbers, then

\begin{displaymath}
\begin{split}
&U_{{P^{\mathrm{**}}}_T\mathrm{\circ }P_T}\left({\pi' \mathrm{\textrm{´}}}_{\mathrm{\Omega }}\mathrm{\circ }{P^{\mathrm{**}}}_T\mathrm{\circ }P_T\mathrm{\circ }0^{\left|U\left(P_{\sum{}}\mathrm{\circ }{P^{\mathrm{**}}}_T\mathrm{\circ }P_T\mathrm{\circ }N\right)\mathrm{*}\mathrm{0\dots 0}\right|}\mathrm{1}\mathrm{\circ }U\left(P_{\sum{}}\mathrm{\circ }{P^{\mathrm{**}}}_T\mathrm{\circ }P_T\mathrm{\circ }N\right)\mathrm{*}\mathrm{0\dots 0}\right)\mathrm{=} \\
&U\left({\pi' \mathrm{\textrm{´}}}_{\mathrm{\Omega }}\mathrm{\circ }{P^{\mathrm{**}}}_T\mathrm{\circ }P_T\mathrm{\circ }0^{\left|U\left(P_{\sum{}}\mathrm{\circ }{P^{\mathrm{**}}}_T\mathrm{\circ }P_T\mathrm{\circ }N\right)\mathrm{*}\mathrm{0\dots 0}\right|}\mathrm{1}\mathrm{\circ }U\left(P_{\sum{}}\mathrm{\circ }{P^{\mathrm{**}}}_T\mathrm{\circ }P_T\mathrm{\circ }N\right)\mathrm{*}\mathrm{0\dots 0}\right)\mathrm{=} \\
&U\left({\pi' \mathrm{\textrm{´}}}_{\mathrm{\Omega }}\mathrm{\circ }{P^{\mathrm{**}}}_T\mathrm{\circ }P_T\mathrm{\circ }0^{\left|U\left(P_{\mathrm{\sum }}\mathrm{\circ }{P^{\mathrm{**}}}_T\mathrm{\circ }P_T\mathrm{\circ }N\right)\right|}\mathrm{1}\mathrm{\circ }U\left(P_{\mathrm{\sum }}\mathrm{\circ }{P^{\mathrm{**}}}_T\mathrm{\circ }P_T\mathrm{\circ }N\right)\right)\mathrm{=}\\
&{{BB}^{\mathrm{+}}}_{{P^{\mathrm{**}}}_T\mathrm{\circ }P_T}\mathrm{(}N\mathrm{)\ }
\end{split}
\end{displaymath}

\noindent Therefore, we obtain our intended program in the form ${\pi' \textrm{´}}_{\mathrm{\Omega }}\circ {P^{**}}_T\circ P_T\circ 0^{\left|\rho \right|}1\circ \rho $ that calculates ${{BB}^+}_{{P^{**}}_T\circ P_T}(N)$ running on submachine $U_{{P^{**}}_T\circ P_T}$. \\\\

For the second part of the theorem, we have that, from the definition of language $L$ and since ${\pi' \mathrm{\textrm{´}}}_{\mathrm{\Omega }}\mathrm{\circ }{P^{\mathrm{**}}}_T\mathrm{\circ }P_T$ has a finite size $\ge 1$, then, for every $N$,

\[\left|U\left(P_{\mathrm{\sum }}\mathrm{\circ }{P^{\mathrm{**}}}_T\mathrm{\circ }P_T\mathrm{\circ }N\right)\right|\le N+1\] 

\noindent So, 

\[\left|{\pi' \mathrm{\textrm{´}}}_{\mathrm{\Omega }}\mathrm{\circ }{P^{\mathrm{**}}}_T\mathrm{\circ }P_T\mathrm{\circ }0^{\left|U\left(P_{\sum{}}\mathrm{\circ }{P^{\mathrm{**}}}_T\mathrm{\circ }P_T\mathrm{\circ }N\right)\right|}\mathrm{1}\mathrm{\circ }U\left(P_{\sum{}}\mathrm{\circ }{P^{\mathrm{**}}}_T\mathrm{\circ }P_T\mathrm{\circ }N\right)\right|\le 2N+C\] 

We will also have that, if 

\[\left|{\pi' \mathrm{\textrm{´}}}_{\mathrm{\Omega }}\mathrm{\circ }{P^{\mathrm{**}}}_T\mathrm{\circ }P_T\mathrm{\circ }0^{\left|U\left(P_{\sum{}}\mathrm{\circ }{P^{\mathrm{**}}}_T\mathrm{\circ }P_T\mathrm{\circ }N\right)\right|}\mathrm{1}\mathrm{\circ }U\left(P_{\sum{}}\mathrm{\circ }{P^{\mathrm{**}}}_T\mathrm{\circ }P_T\mathrm{\circ }N\right)\right|-1<N \]

\noindent and $\mathrm{0\dots 0\ }$ is minimal, then

\[\left|{\pi' \mathrm{\textrm{´}}}_{\mathrm{\Omega }}\mathrm{\circ }{P^{\mathrm{**}}}_T\mathrm{\circ }P_T\mathrm{\circ }0^{\left|U\left(P_{\sum{}}\mathrm{\circ }{P^{\mathrm{**}}}_T\mathrm{\circ }P_T\mathrm{\circ }N\right)\mathrm{*}\mathrm{0\dots 0}\right|}\mathrm{1}\mathrm{\circ }U\left(P_{\sum{}}\mathrm{\circ }{P^{\mathrm{**}}}_T\mathrm{\circ }P_T\mathrm{\circ }N\right)\mathrm{*}\mathrm{0\dots 0}\right|\mathrm{\le }N+2\] 

Thus, it leads us to the conclusion that there is a constant $C$ such that

\[\left|{\pi' \mathrm{\textrm{´}}}_{\mathrm{\Omega }}\mathrm{\circ }{P^{\mathrm{**}}}_T\mathrm{\circ }P_T\mathrm{\circ }0^{\left|U\left(P_{\sum{}}\mathrm{\circ }{P^{\mathrm{**}}}_T\mathrm{\circ }P_T\mathrm{\circ }N\right)\mathrm{*}\mathrm{0\dots 0}\right|}\mathrm{1}\mathrm{\circ }U\left(P_{\sum{}}\mathrm{\circ }{P^{\mathrm{**}}}_T\mathrm{\circ }P_T\mathrm{\circ }N\right)\mathrm{*}\mathrm{0\dots 0}\right|\le 2N+C \]

\end{pf}

\end{thm}

\section[Sub-uncomputability]{Sub-Uncomputability: recursive relative uncomputability}

Now we will prove the crucial, yet simple, result that underpins this paper. 

Let ${P'\textrm{´}}_T$ be a total function and $U_{{P'}_T}$ a Turing submachine. Then, we prove in theorem \ref{thm11.1} that function ${{BB}^+}_{{P'}_T}\left(N\right)$ is relatively uncomputable\textbf{\textit{ }}by any program on submachine $U_{{P'\textrm{´}}_T}$. Or: there is no subprogram that, for every input $N$, returns an  output equal to ${{BB}^+}_{{P'\textrm{´}}_T}\left(N\right)$. Once ${{BB}^+}_{{P'\textrm{´}}_T}\left(N\right)$ is computable, then we say it is \textbf{sub-uncomputable}. 

Note that, since $U_{{P'}_T}$ is a time-bounded Turing machine,  ${{BB}^+}_{{P'\textrm{´}}_T}$ promptly defines a time hierarchy \cite{Papadimitriou1994, Arora2009} in computational complexity theory. In fact, one can define\textbf{ }${BB}^+$\textbf{ }upon submachines that are not necessarily time-bounded Turing machines, so that, on the other hand, ${{BB}^+}_{{U}/{f}}(N)$ will be a function that defines arbitrary subrecursive hierarchies.

A more intuitive way to understand what is going on is to look for a program and concatenate its input, such as $U_{{P'\textrm{´}}_T}\left(P*N\right)\ $for instance. Where ``$*$'' denotes the optimal functionalizing concatenation, and not necessarily the ``concatenation'' ``$\circ $'' defined. See section \ref{introlanguage}. In fact, this applies to any way to compress the information of $P$ and $N$ in an arbitrary subprogram. Anyhow, the theorem \ref{thm11.1} will be proved for both ``$\circ $'' and for ``$*$''. 

We avail ourselves of the same idea used in the demonstration of Chaitin's incompleteness theorem \cite{Chaitin1971}. Now, however, to demonstrate an uncomputability relative to the submachine $U_{{P'\textrm{´}}_T}$.

When $N$ is given as input to any program $P$, it comes in its compressed form with size $\cong H\left(N\right)$. In fact, we use the property 

\begin{displaymath}
\left|P\mathrm{\circ }N\right|\mathrm{\le }C\mathrm{+|}P\mathrm{|+}C'\mathrm{\textrm{´}+}{{log}_{\mathrm{2}} N\ }\mathrm{+(1+}\epsilon \mathrm{)}{{log}_{\mathrm{2}} \mathrm{(}{{log}_{\mathrm{2}} N\ }\mathrm{)}\ }
\end{displaymath}

\noindent Whereby,

\[\left|P\circ N\right|\cong C+H\left(N\right) \]

But, as already known by the AIT, for any constant $\mathrm{C}$\textbf{\textit{ }}there is a big enough $N_0\ $such that\textbf{\textit{ }}$C+H\left(N_0\right)<N_0$. Therefore, according to\textbf{\textit{ }}the definition of ${{BB}^+}_{{P'\textrm{´}}_T}$, the output of $P\circ N_0$ when run on submachine $U_{{P'\textrm{´}}_T}$, will be taken into account when one calculates ${{BB}^+}_{{P'\textrm{´}}_T}\left(N_0\mathrm{\ }\right)$. Thus, necessarily,

\[{{BB}^{\mathrm{+}}}_{{P'\mathrm{\textrm{´}}}_T}\left(N_0\mathrm{\ }\right)\mathrm{\ge }U_{{P'\mathrm{\textrm{´}}}_T}\left(P\mathrm{\circ }N_0\right)\mathrm{+1>}U_{{P'\mathrm{\textrm{´}}}_T}\left(P\mathrm{\circ }N_0\right)\] 

\noindent Which will lead to contradiction, if $P$ computes ${{BB}^+}_{{P'\textrm{´}}_T}$ when running on submachine $U_{{P'\textrm{´}}_T}$. The same holds for ``$*$''.

Also, following the same argument, it can be shown promptly that ${{BB}^+}_{{P'\textrm{´}}_T}$ is a \textbf{relatively incompressible}, or\textbf{ sub-incompressible, }function by any subprogram smaller than or equal to $N$. That is, no program of size $\le N$ running on\textbf{\textit{ }}$U_{{P'\mathrm{\textrm{´}}}_T}$ will result in an output larger than or equal to ${{BB}^+}_{{P'\textrm{´}}_T}\left(N\right)$. This follows directly from the definition of the function ${{BB}^{\mathrm{+}}}_{{P'\mathrm{\textrm{´}}}_T}$. Besides, if one proves sub-incompressibility of ${{BB}^+}_{{P'\textrm{´}}_T}\left(N\right)$ at first hand, then sub-uncomputability follows as a corollary.

Note that nothing can be said as yet about the relative \textnormal{randomness (or
incompressibility)} of ${{\Omega}}_{{P^{**}}_T\circ P_T}$, which we aim indeed to study in further work.

\begin{thm} \label{thm11.1}
\textnormal{\newline}
Let $N$ be an arbitrary natural number.
Let ${P'\textrm{´}}_{T\ }$be a total function.
Let $U_{{P'\textrm{´}}_T}\ $be a Turing submachine.
Then, function ${{BB}^+}_{{P'\textrm{´}}_T}\left(N\right)$ is uncomputable\textbf{\textit{ }}by any program on $U_{{P'\textrm{´}}_T}$.
That is, one can prove a strict dominance:

\[\mathrm{\forall }P\mathrm{\exists }N_0\mathrm{\forall }N\mathrm{(\ }N\mathrm{\ge }N_0\mathrm{\ }\mathrm{\to }\mathrm{\ }U_{{P'\mathrm{\textrm{´}}}_T}\left(P\mathrm{*}N\right)\mathrm{<}{{BB}^{\mathrm{+}}}_{{P'\mathrm{\textrm{´}}}_T}\left(N\right)\mathrm{)}\] 

\noindent and

\noindent 
\[\mathrm{\forall }P\mathrm{\exists }N_0\mathrm{\forall }N\mathrm{(\ }N\mathrm{\ge }N_0\mathrm{\ }\mathrm{\to }\mathrm{\ }U_{{P'\mathrm{\textrm{´}}}_T}\left(P\mathrm{\circ }N\right)\mathrm{<}{{BB}^{\mathrm{+}}}_{{P'\mathrm{\textrm{´}}}_T}\left(N\right)\mathrm{)}\] 

\begin{pf}
\textnormal{\newline}

Take an arbitrary program $P$ in the language $L$.
From the definition of language $L$ and of $U$, we have that:

\[\left|P\mathrm{\circ }N\right|\mathrm{\le }C\mathrm{+|}P\mathrm{|+}C'\mathrm{\textrm{´}+}{{\mathrm{log}}_{\mathrm{2}} N\ }\mathrm{+(1+}\epsilon \mathrm{)}{{\mathrm{log}}_{\mathrm{2}} \mathrm{(}{{\mathrm{log}}_{\mathrm{2}} N\ }\mathrm{)}\ }\] 

We have, from the definition of $P*N$ in \ref{introlanguage}, that $|P*N|\le |P\circ N|$.  

Let $C\mathrm{+}\left|P\right|\mathrm{+}C'\mathrm{\textrm{´}=}C''\mathrm{\textrm{´}\textrm{´}}$.

It is well-known that:

\[\mathrm{\exists }N_0\mathrm{\forall }N\mathrm{(\ }N\mathrm{\ge }N_0\mathrm{\ }\mathrm{\to }\mathrm{\ }C''\mathrm{\textrm{´}\textrm{´}+}{{\mathrm{log}}_{\mathrm{2}} N\ }\mathrm{+(1+}\epsilon \mathrm{)}{{\mathrm{log}}_{\mathrm{2}} \mathrm{(}{{\mathrm{log}}_{\mathrm{2}} N\ }\mathrm{)}\ }\mathrm{<}N\mathrm{)}\] 

Thus, 

\[\mathrm{\exists }N_0\mathrm{\forall }N\mathrm{(}N\mathrm{\ge }N_0\mathrm{\ }\mathrm{\to }\mathrm{\ |}P\mathrm{*}N\mathrm{|}\mathrm{\le }\left|P\mathrm{\circ }N\right|\mathrm{<}N\mathrm{)}\]

Consequently, from the definition of\textbf{\textit{ }}${{BB}^+}_{{P'\textrm{´}}_T}$, we will have that there is $N_0$ such that

\[\mathrm{\forall }N\left(\mathrm{\ }N\mathrm{\ge }N_0\mathrm{\ }\mathrm{\to }\mathrm{\ }U_{{P'\mathrm{\textrm{´}}}_T}\left(P\mathrm{\circ }N\right)\mathrm{<}U_{{P'\mathrm{\textrm{´}}}_T}\left(P\mathrm{\circ }N\right)\mathrm{+1}\mathrm{\le }{{BB}^{\mathrm{+}}}_{{P'\mathrm{\textrm{´}}}_T}\left(N\right)\right)\]

\noindent and

\[\mathrm{\forall }N\left(\mathrm{\ }N\mathrm{\ge }N_0\mathrm{\ }\mathrm{\to }\mathrm{\ }U_{{P'\mathrm{\textrm{´}}}_T}\left(P\mathrm{*}N\right)\mathrm{<}U_{{P'\mathrm{\textrm{´}}}_T}\left(P\mathrm{\circ }N\right)\mathrm{+1}\mathrm{\le }{{BB}^{\mathrm{+}}}_{{P'\mathrm{\textrm{´}}}_T}\left(N\right)\mathrm{\ }\right)\] 

\end{pf}

\end{thm}

\section[Conclusion]{Conclusions}

We have defined Turing submachines as another terminology for total Turing machines, but emphasizing the property that there are always non-reducibly more powerful Turing machines from which total Turing machines are subsystems \cite{Abrahao2016}. 

Assuming $P_T$ as program that computes a total function, a Turing submachine $U_{{P^{**}}_T\circ P_T}$ is built such that $U_{{P^{**}}_T\circ P_T}$ is a proper extension of $U_{P_T}$, by proving that ${P^{**}}_T\circ P_T$ is also a program that computes a total function. Furthermore, a relative and computable halting probability ${{\Omega}}_{{P^{**}}_T\circ P_T}$ --- in the case, a time-bounded one --- was defined  regarding which was proved that there are finite lower approximations $\rho $ that can be used by a program $P$ when running on $U_{{P^{**}}_T\circ P_T}$ to compute values of ${{BB}^+}_{{P^{**}}_T\circ P_T}\left(N\right)$, so that  $2N+C\ge |P\circ \rho |\ge N+1$, where $C$ is a constant. The submachine $U_{{P^{**}}_T\circ P_T}$ is a time-bounded Turing machine that increases the computation power of the time-bounded Turing machine limited by $P_T$. And it does that by recursively allowing some programs to calculate values of ${{BB}^+}_{{P^{**}}_T\circ P_T}\left(N\right)$ for lower approximations to ${{\Omega}}_{{P^{**}}_T\circ P_T}$ through a self-diagonalizing procedure.

We also proved that, for every Turing submachine $U_{{P\mathrm{'}}_T}$, if ${P'\mathrm{\textrm{´}}}_T$ is a program that computes a total function, then the computable function ${{BB}^+}_{{P'}_T}\left(N\right)$ is relatively uncomputable by any program running on $U_{{P\mathrm{'}}_T}$ --- in the ``same manner'' that the Busy Beaver function $BB'\textrm{´}(N)$ is in relation to any program on $U$. Also, by the very definition of ${{BB}^+}_{{P'}_T}$, there cannot be any program of size $\le N$ running  on $U_{{P\mathrm{'}}_T}$ that generates an output higher than or equal to ${{BB}^+}_{{P'}_T}\left(N\right)$. In other words, we have shown in the last theorem \ref{thm11.1} that relative uncomputability and relative algorithmic incompressibility are ubiquitous phenomena for submachines. Thus, a more powerful containing system, from which a subsystem is a submachine, can promptly computes a problem of irreducible complexity compared to this subsystem. These results allow us to build a hierarchy of subrecursive classes built on relative uncomputability and relative incompressibility of the Busy Beaver function.

Moreover, since we have proved that ${P^{**}}_T\circ P_T$ is total, function ${{BB}^+}_{{P^{**}}_T\circ P_T}$ becomes sub-uncomputable by any program running on $U_{{P^{**}}_T\circ P_T}$, but still reachable by giving lower approximations to ${{\Omega}}_{{P^{**}}_T\circ P_T}$ as inputs to (at least) one kind of program running on $U_{{P^{**}}_T\circ P_T}$. Also, function ${{BB}^+}_{{P^{**}}_T\circ P_T}\left(N\right)$ is always incompressible by any program of size $\le N$ running on $U_{{P^{**}}_T\circ P_T}$. This was designed to mimic what a universal Turing machine can do with lower approximations to ${\Omega}$ with the purpose of calculating values of $BB'\textrm{´}(N)$ while maintaining this function as a ``yardstick'' of irreducible algorithmic complexity/information  --- see \cite{Chaitin2013, Chaitin2012, Chaitin2014}. 

Note that this does not mean that ${{\Omega}}_{{P^{**}}_T\circ P_T}$ is random/incompressible in relation to $U_{{P^{**}}_T\circ P_T}$, whereas it holds for relative halting probabilities across the Turing degrees for example \cite{Downey2005}. In fact, not ``all uncomputabilities'' of a first order hypercomputer were relativized in relation\textbf{\textit{ }}to a universal Turing machine. We propose for further research to investigate to what extent one can make a Turing machine be made to ``behave'', in relation to one of its submachines (or subsystems), as if it were a hypercomputer. Analogously, one can also ask if there is such a thing as ``sub-randomness'' of a bit string or a real number, in the same way that there is sub-uncomputability of a function. For example, these inquiries might encompass studying both the computational complexity of running such proper subrecursive extensions and its relations to Rice's theorem and to Turing degrees within computability theory \cite{Kozen1980, Hoyrup2016}. 

Alongside with the pursuit for such limiting results, this kind of recursive relativization has already shown to be useful in order to build theoretical models of computable systems. For example, we have defined in \cite{Abrahao2015} evolutionary\textbf{\textit{ }}models without the need of a hypercomputable environment (or ``Nature'') that are fully analogous to Chaitin's\textbf{\textit{ }}hypercomputable\textbf{\textit{ }}models of the open-ended cumulative (darwinian) evolution of software. That is, we have proved that the open-ended evolution of subprograms can be as fast as the open-ended evolution of programs, through simulating the uncomputability and incompressibility of the Busy Beaver function for Turing submachines.
\\	
\\

\begin{flushleft}

\bibliography{References}

\end{flushleft}

\end{document}